\documentclass{aastex63}
\usepackage{epsf,amssymb,amsmath}
\usepackage{graphicx}
\def\gaia{{\textit{Gaia}}}
\def\simbad{SIMBAD}
\def\ruwe{{\texttt{ruwe}}}

\def\catlongname{{\gaia} Nearby Accelerating Star Catalog}
\def\catshortname{GNASC}
\def\au{au}
\def\Msolar{\ifmmode {\text{M}_{\odot}}\else {$\text{M}_\odot$}\fi}

\def\ncatstars{29,684} %29609

\begin{document}

\title{A catalog of nearby accelerating star candidates in \gaia\ DR3}
\vskip 7ex

\author[0000-0002-6381-1615]{Marc L. Whiting}
\affil{Department of Physics \& Astronomy, University of Utah, Salt Lake City, UT 84112}
\email{e-mail: mwhiting@astro.utah.edu}

\author[0000-0001-9764-7385]{Joshua B. Hill}
\affil{Department of Physics \& Astronomy, University of Utah, Salt Lake City, UT 84112}
\email{e-mail: joshua.hill@astro.utah.edu}

\author[0000-0001-7558-343X]{Benjamin C. Bromley}
\affil{Department of Physics \& Astronomy, University of Utah, Salt Lake City, UT 84112}
%\email{e-mail: bromley@physics.utah.edu}

\author[0000-0003-0214-609X]{Scott J. Kenyon}
\affil{Smithsonian Astrophysical Observatory,
60 Garden Street, Cambridge, MA 02138}
%\email{e-mail: skenyon@cfa.harvard.edu}

\shortauthors{Whiting \& Hill et al.}

\begin{abstract}
We describe a new catalog of accelerating star candidates with Gaia $G \le$ 17.5 mag and distances $d \le$ 100~pc. Designated as \catlongname\ (\catshortname), it contains \ncatstars\ members identified using a supervised machine-learning algorithm trained on the Hipparcos--Gaia Catalog of Accelerations (HGCA), \gaia\ Data Release~2, and \gaia\ Early Data Release~3. We take advantage of the difference in observation timelines of the two \gaia\ catalogs and information about the quality of the astrometric modeling based on the premise that acceleration will correlate with astrometric uncertainties. Catalog membership is based on whether constant proper motion over three decades can be ruled out at high confidence (greater than 99.9\%). Test data suggest that catalog members each have a 68\%\ likelihood of true astrometric acceleration; subsets of the catalog perform even better, with the likelihood exceeding 85\%. We compare the \catshortname\ with \gaia\ Data Release 3 and its table of stars for which acceleration is detected at high confidence based on precise astrometric fits. Our catalog, derived without this information, captured over 96\% of sources in the table that meet our selection criteria. In addition, the \catshortname\ contains bright, nearby candidates that were not in the original Hipparcos survey, including members of known binary systems as well as stars with companions yet to be identified. It thus extends the HGCA and demonstrates the potential of the machine-learning approach to discover hidden partners of nearby stars in future astrometric surveys.

\end{abstract}

\section{Introduction}

For nearly 180~years, irregular motions of stars and planets have fueled new discoveries. Bessel's analysis of Sirius and Procyon led to the detections of Sirius b and Procyon b \citep{bessel1844,bond1862,struve1873}. Deviations in the orbit of Uranus enabled the discovery of Neptune \citep{galle1846}. For several decades, van de Kamp used astrometric measurements of Barnard's star to place limits on possible planetary companions \citep[e.g.,][]{vandekamp1956,vandekamp1969};  van de Kamp's detection of irregular motion was at the limit of his instrumentation and has never been confirmed \citep[e.g.,][]{ribas2018, lubin2021}.

The launch of the Hipparcos and Gaia satellites ushered in a new age of high-precision astrometric measurements that enabled the detection of unseen stellar and planetary companions to nearby stars \citep{hipparcos1997, gaia2016}. The new \gaia~Data Release~3 catalog \citep[DR3;][]{gaia2022} is a culmination. It includes a table of astrometric acceleration solutions for almost 340,000 sources \citep[\texttt{nss\_acceleration\_astro};][]{halbwachs2022, holl2022}, whose proper motion has significantly changed over Gaia's first 34-months of observations. Each star on this list has a gravitating companion, whether known \citep{brandtm2021} or a potential new discovery \citep[e.g.][]{kervella2022, pearce2022}.
 
Together, the Hipparchos and Gaia catalogs sample the proper motion of nearby stars at times separated by decades. \citet{brandt2018} took advantage of this long temporal baseline to create the Hipparcos-Gaia Catalog of Accelerations (HGCA), derived from differences in proper motion between Hipparcos and \gaia~Data Release~2 \citep[DR2][]{gaia2018}. The HGCA, and its revised version based on \gaia\ Early Data Release~3 (EDR3) data \citep{gaia2021}, has placed constraints on the properties of known low-luminosity companions \citep{brandt2019}, and led to the discovery of substellar partners \citep[e.g.,][]{currie2020, kuzuhara2022}.

The HGCA has acceleration solutions for 115,346 nearby stars, drawn from the 117,955 sources with astrometry from Hipparcos \citep{perry1997}. These stars were selected from the Hipparcos Input Catalog \citep[e.g.,][]{turon1995}, which is limited in $V$-band photometry to 12.7~mag but is complete across the whole sky only to $V=7.3$. The HGCA thus missed a pool of nearby potential accelerating star candidates. The DR3 \texttt{nss\_acceleration\_astro} table may also miss promising candidates, as it lists only sources with high-significance acceleration solutions over the shorter time frame of Gaia observations. 

A machine-learning approach offers the possibility of filling in the gap between the HGCA and DR3. Following the ``ML'' paradigm for supervised learning, HGCA sources serve as a training set with known accelerations. Each of these sources has ``features'', in this case, astrometric measurements, quality of fit estimators, and stellar properties from Gaia DR2 and EDR3, that may serve as a basis for predicting the HGCA acceleration. For example, the EDR3 model parameter \ruwe\ --- the renormalized unit weighted error --- is a measure of how well a constant proper motion solution fits each source. When a star is measurably accelerating, this indicator is high compared to sources that are just drifting in the plane of the sky \citep{brandt2018, belokurov2020, kervella2022, pearce2022}. With a broad range of features common to nearly all EDR3 sources, we can train a ML algorithm to predict whether stars are likely to be accelerating, even when they are not members of the HCGA.

A challenge for this application of machine learning is a mismatch in the sensitivity to nonlinear motion in the HGCA and \gaia\ DR3. The HGCA's long time baseline is optimal for detecting orbital periods of decades, while in \gaia\ DR3, astrometric measurements select most efficiently for more rapid orbits with periods of several years. The risk in applying the ML code to \gaia\ sources when it has been trained with the HGCA is that there may not be enough information available in the \gaia\ data to identify sources with longer-period orbits. For example, the Gaia \ruwe\ parameter may be low, even consistent with linear drift, if a source is accelerating slowly on a longer-period orbit \citep[e.g.,][]{pearce2022, kervella2022}. A goal here is to explore whether a combination of \gaia\ astrometric features, including \ruwe\ but also the detailed errors in parallax and proper motion, along with the changes in astrometry from DR2 to EDR3, might reveal promising candidates with significant acceleration from a nearby stellar or substellar companion.

In this work, we use a machine-learning algorithm to generate a new catalog of bright, accelerating star candidates at distances within 100~pc of the Sun. To describe this approach and to introduce our catalog, we organize this paper as follows: In \S2\, we explain the initial steps and selection of stars in the HGCA. The machine-learning algorithm is described, including how we train and test for stellar accelerations. In \S3, the \catlongname (\catshortname), derived from the machine-learning predictor, is then presented. In \S 4, we discuss our findings.

\section{Method: Background and Machine Learning}\label{sec:method}

Measuring a star's change in proper motion over a long time can reveal if it has a hidden companion or is bound to an observed neighboring star. To explore the sensitivity of this approach to finding a star's companion, we estimate the physical velocity of the star, as projected on the plane of the sky, in terms of its proper motion, $\vec{\mu}$ and parallax
$\varpi$:
\begin{equation}
    \vec{v}_\perp \approx 
    4.7074 \left[\frac{\vec{\mu}}{\text{mas/yr}}\right] \left[\frac{\varpi}{\text{mas}}\right]^{-1} \ {\rm km~s^{-1}}. 
\end{equation}
If that physical velocity is associated with the star's orbital motion with a bound companion, then its orbital speed is roughly
\begin{equation}
    v \sim \left[\frac{G M_\bullet}{a}\right]^{1/2},
\end{equation}
where $M_\bullet$ is the mass of the star's companion, $a$ is the physical separation between the two, and $G$ is the gravitational constant. We anticipate approximately linear acceleration if the pair's orbital period is long. Thus, the magnitude of the change in velocity has the form
\begin{equation}
    \frac{\Delta v}{\Delta t} \sim \frac{G M_\bullet}{a^2},
\end{equation}
where $\Delta t$ is the temporal baseline of the observations, roughly 25 years for the HGCA. For linear acceleration to dominate the velocity change, we consider orbital configurations with $r \gtrsim 30$~\au\ when the mass of the star is equivalent to that of the Sun. Then, in terms of proper motion, 
\begin{equation}\label{eq:dmu}
    \Delta \mu\sim 2.5 
    \left[\frac{\varpi}{1~\text{mas}}\right]
    \left[\frac{M_\bullet}{1~\Msolar}\right]
    \left[\frac{a}{20~\text{\au}}\right]^{-2} 
    \left[\frac{\Delta t}{25~\text{yr}}\right]
    \ \text{mas/yr}.
\end{equation}
This expression allows us to connect a measured constant acceleration with a star's orbital configuration. For example, at a distance of 10 pc, the star's reflex motion from a 0.1~\Msolar\ companion at 30~\au\ is  $\Delta\mu \approx 10$~mas/yr. At that same parallax, a star with a 10~\Msolar\ companion would have a similar change in proper motion at a separation of 300~\au. The pair separation, in this case, would be around 0.5~arcminutes. Our goal is not to explicitly examine these changes in proper motion but to identify promising candidate accelerating stars in a single catalog based on machine learning and training with Hipparcos and Gaia, as we describe next. Equation~(\ref{eq:dmu}) provides a guide to the mass and location of companions of these candidate stars. 

\subsection{Machine learning on known accelerating stars}\label{subsec:ml}

The original HGCA introduced by \citet{brandt2018} incorporated 98\% of the Hipparchos survey and \gaia~DR2 cross-matches to assess whether each source was consistent with linear drift over the time between the two surveys. In EDR3, the astrometric uncertainty level was greatly reduced compared to DR2, especially for bright stars \cite{gaia2021}. Thus, \citet{brandt2021} updated the HGCA using EDR3 data to generate a table of 115,346 stars, each with an EDR3 \texttt{source\_id}. This number is a small fraction of the  millions of \gaia\ sources with brightness comparable to those in the Hipparchos catalog, or even the number of stars within 100~pc of the Sun. Thus, there is potential to use a machine learning approach to extend the results of the HGCA to predict stellar accelerations more broadly in the \gaia\ data. 

The principle result of the HGCA is assessing whether each star is consistent with linear drift over several decades. The metric is a $\chi^2$ value, corresponding to the difference between a linear drift model and the reported proper motions in the Hipparcos catalog and EDR3. The linear drift model has two parameters ($m=2$), a constant drift speed in each of the two components of proper motion in the plane of the sky, while the data from the two surveys provide four observations (proper motion in ra and dec over each of the surveys; $N=4)$. The number of degrees of freedom in the model fit is $dof = N-m =2$. Assuming standard errors, we can associate each $\chi^2$ value with a $p$-value, giving the likelihood that the star's motion is consistent with the hypothesis of pure, linear drift. For example, a star with $\chi^2$ of 11.8 corresponds to $p$-value of 99.7\%, indicating a ``3-$\sigma$'' outlier \citep{brandt2021}. Smaller $\chi^2$ values mean a better fit between model and observations, and a larger $p$-value and greater likelihood of linear drift. Larger $\chi^2$ values mean a worse fit and a smaller likelihood that a star drifts linearly. At a value of $\chi^2 = 28.75$, corresponding to a ``5-$\sigma$'' result, the $p$-value is $5.7\times 10^{-7}$. Even in the full \gaia\ catalog, we would not expect to find any star drifting freely to have such a small $p$-value due to random errors. Conversely, if astrometric errors are accurate, every star with $\chi^2 \geq 28.75$ is expected to be experiencing some acceleration.

We aim to set up a machine-learning predictor of stellar acceleration, trained with the HGCA, that can generally be applied to stars in the \gaia\ survey. To do so, we primarily have two decisions related to the learning algorithm's input parameters and the predictor's output.  We could work with the \gaia~EDR3 data model for the input, selecting a subset of features (as below), including proper motion and astrometric error indicators. As it turns out, it is also advantageous to include some proper motion information from DR2. We confirm this advantage and settle on the final list of features to have in the machine learning by training and testing on subsets of HGCA stars. 

To obtain source identifiers of objects in both \gaia~DR2 and EDR3, we use the \texttt{astroquery.py} to query \gaia's \texttt{gaiaedr3.dr2\_neighbourhood} table \citep{astropy:2018}. Cross matches are accepted if differences in the \texttt{photo\_g\_mean\_mag} are less than 1~mag. When multiple DR2 \citet{2016A&A...595A...1G, gaia2018} sources are returned in the neighborhood of an EDR3 source, we choose the closest DR2 source in angular separation. If that DR2 source does not satisfy the photometry cut, no DR2 source is matched for that EDR3 star. The outcome is a list of 115,160 HGCA stars associated with EDR3 and DR2 sources, from which we draw training and test sets for machine learning.

\begin{figure}
%\centerline{
\centerline{
\includegraphics[width=6.5in]{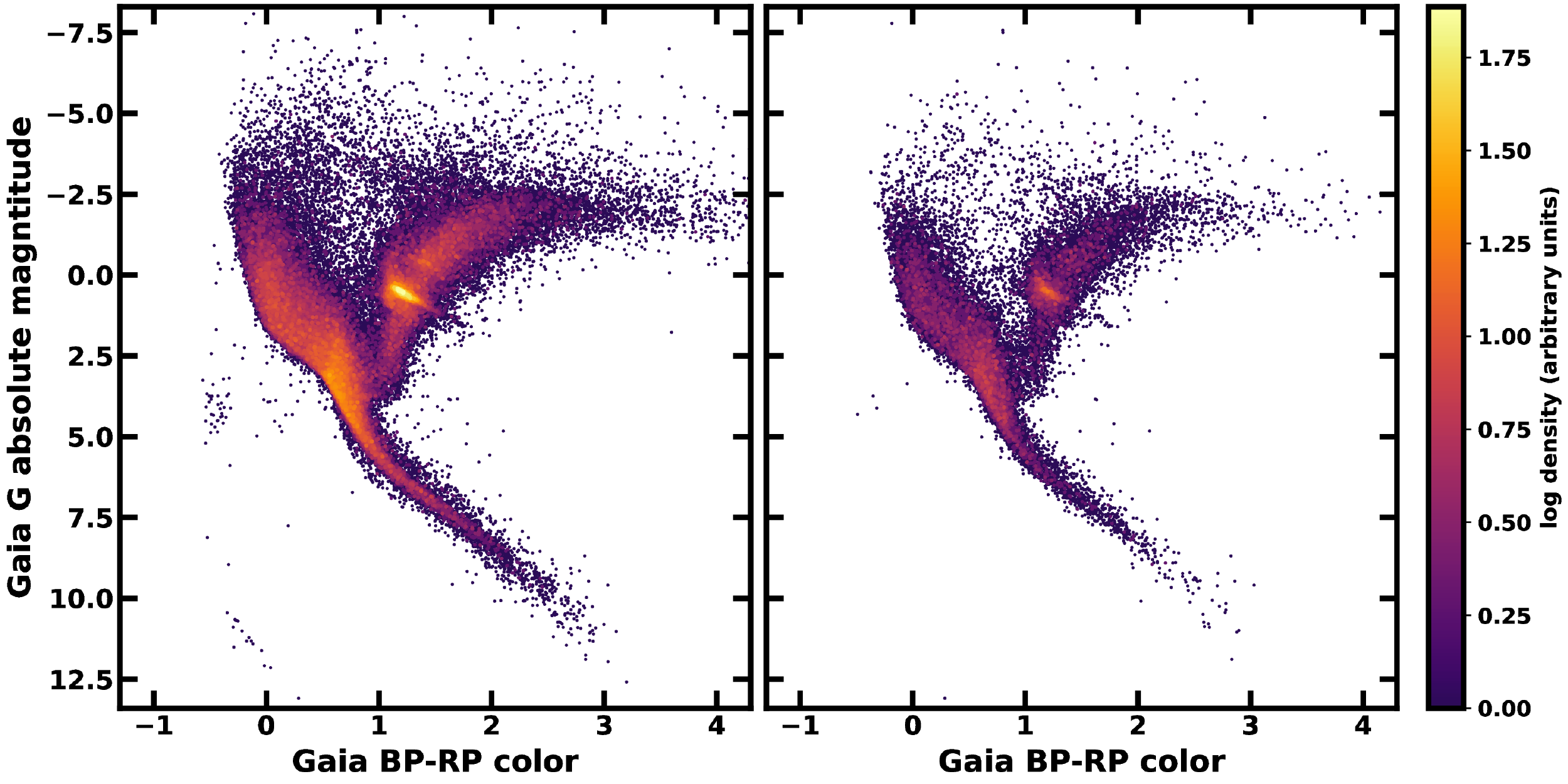} % uodated using -density 300 in convert...2023/01/20
}
%\vspace{-1.5in}
   \caption{The color-magnitude diagram (CMD) from the Hipparcos-Gaia Catalog of Accelerations.
    The left panel is the full catalog from \citet{brandt2021} with 115,346 stars. As described in the text, we searched for counterparts in the Gaia DR2 catalog and found matches for all but 186 sources. The right panel shows the CMD for the reduced HGCA catalog containing only the accelerating stars with $\chi^2 \geq~ 28.75$, which have cross matches in DR2 and EDR3.  The distributions of spectral types within the full catalog and the subset of accelerating stars are similar; however, no significant acceleration is attributed to white dwarfs (lower left in the left CMD, near [11, -0.2]). Given that roughly a third of the stars show motion inconsistent with linear drift, there may also be a dearth of accelerating hot subdwarfs (the cluster of stars near [4,-0.3] in the right CMD as compared to the same region in the left panel).
    \label{fig:hi}}
\end{figure}

Figure~\ref{fig:hi} shows the color-magnitude diagram (CMD) of stars in the HGCA, alongside ``accelerating'' stars for which linear drift can be formally ruled out with 99.3\% confidence or greater ($\chi^2 \geq 11.8$; see \citealt{brandt2021}). No difference stands out between the accelerating population and the full HGCA, except that no white dwarfs have detectable acceleration and only a few hot subdwarfs (near [4,-0.3] in the CMD) have a motion that is inconsistent with linear drift.

\begin{figure}
\centerline{
\includegraphics[width=4.5in]{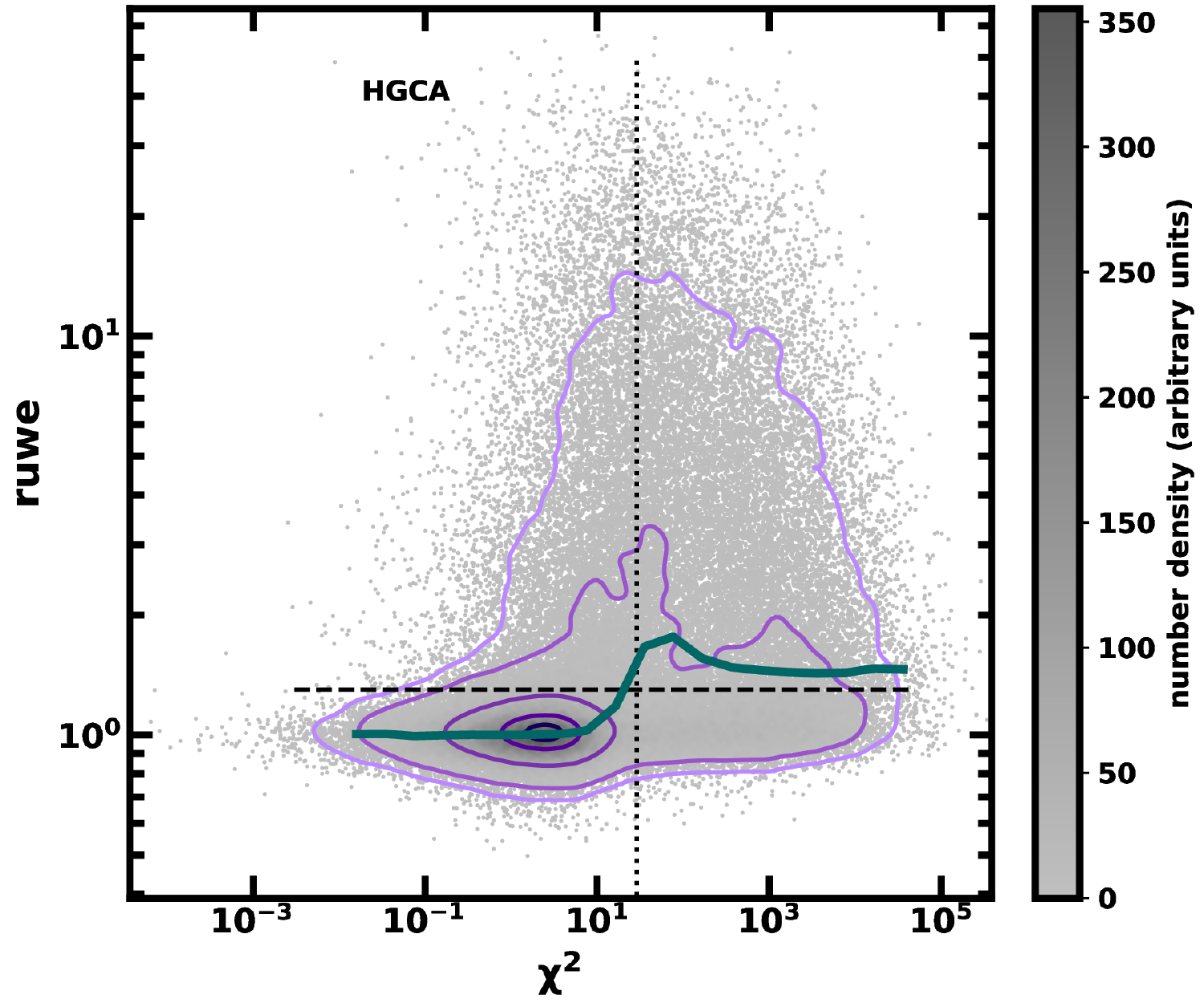} % update with lowere plot density 2023/01/20
}
   \caption{The renormalized unit-weighted error (\texttt{ruwe}) from Gaia EDR3 versus the $\chi^2$ value testing the hypothesis in the HGCA. The horizontal dashed line is at a value of $\text{\texttt{ruwe}}= 1.3$, which nominally delimits binary source (above the line) and single sources (below) in the \gaia\ data. The vertical dotted line is at a value of 28.75, marking 5-$\sigma$ outliers in the distribution of $\chi^2$ values.  
    \label{fig:ruwehgca}}
\end{figure}

Figure \ref{fig:ruwehgca} shows the distribution of \gaia\ EDR3's \ruwe\ parameter as a function of $\chi^2$ derived by \citet{brandt2021}.  Gaia sources with an \ruwe\ above $1.3$, which nominally delimits binary and single stars \citep[e.g.,][]{stassun2021, kervella2022, pearce2022}, have astrometric parameters that are impacted by orbital motion. Many of these stars also show nonlinear motion in the HGCA. Sources with $\chi^2 \ge 28.75$, the 5-$\sigma$ mark in the expected distribution \citep[see][]{brandt2021}, are unambiguously inconsistent with linear drift, and the majority of these stars also have \ruwe\ above the 1.3 threshold. Still, a significant population of stars have low \ruwe, consistent with linear motion in \gaia, yet have accelerations from stellar or substellar companions that become apparent only over the longer HGCA baseline. A goal here is to see if a machine-learning algorithm can identify new candidates of this type.

With a pool of potential features available to us in the HGCA stars in both DR2 and EDR3, we consider what output our machine learning algorithm could provide. The $\chi^2$ values in the HGCA measure how well a star matches the `null hypothesis of the linear drift.' We could set a $\chi^2$ threshold, for example, the 5-$\sigma$ value of 28.75 or a more lenient 11.8, corresponding to a $p$-value of 0.3\%, and assert that any star with a larger $\chi^2$ is labeled `accelerating' while others are `drifting.' These discrete labels could then be the output of a machine-learning \textit{classifier}. Alternatively, we could let the $\chi^2$ value be the predicted output, and work with a machine-learning \textit{regressor} instead. With this choice, we can be more flexible in interpreting the predictions. We follow this plan next.

\subsection{Supervised Learning: Regression}\label{slr}

To obtain an algorithm for predicting a stellar acceleration in \gaia\ data using the HGCA, we begin with a supervised-learning regression that uses test sets with known $\chi^2$ values (the supervised part) based on a suite of features from the \gaia\ EDR3 and DR2 catalogs. These features include \gaia\ EDR3 astrometric model parameters 
(\texttt{pmra}, \texttt{pmra\_error}, \texttt{pmdec}, \texttt{pmdec\_error},  \texttt{parallax}, and \texttt{parallax\_over\_error}), goodness-of-fit measures (\texttt{df.ruwe}, \texttt{astrometric\_excess\_noise\_sig}, and \texttt{astrometric\_gof\_al}), and stellar properties
(\texttt{phot\_g\_mean\_mag} and \texttt{bp\_rp}). To incorporate \gaia\ DR2 information, we also use the difference in proper motion components of DR2 and EDR3 (e.g., $\Delta \mu_\text{RA}$ = \texttt{dr2.pmra} - \texttt{edr3.pmra}). We adjust the DR2 proper motions to correct for a reference frame mismatch \citep{brandt2018, lindegren2020, gaiaastro2021} that impacts sources brighter than $G = 13$~mag \citep[see also][]{cantat-gaudin2021}; following  \citep[Eq.~9 therein]{lindegren2020}, we set
\begin{eqnarray}
\mu_{\alpha\ast} & = & \tilde{\mu}_{\alpha\ast} + \omega_x\cos(\alpha)\sin(\delta)
+\omega_y\sin(\alpha)\sin(\delta) - \omega_z\cos(\delta)\\
\mu_{\delta} & = & \tilde{\mu}_{\delta} - \omega_x\sin(\alpha)
+\omega_y\cos(\alpha),
\end{eqnarray}
where $(\mu_{\alpha\ast},\mu_\delta)$ is the corrected proper motion in terms of vector components in RA ($\alpha$) and dec ($\delta$), while the tilde symbol refers to proper motion in DR2, and a spin vector $(\omega_x,\omega_y,\omega_z) = (-0.077,-0.096,-0.002)$.
Finally, our features also include the magnitude of proper motion in \gaia~EDR3,
\begin{equation}\label{tpm}
\mu_{tot} = \sqrt{\mu_{\alpha\ast}^2 + \mu_{\delta}^2}
\end{equation}
and that of the error vector (also adding errors in quadrature), as well as the magnitude of the speed difference between the two catalogs. 

Our final parameter space of features thus has 16 dimensions, albeit with some redundant information. The set of features is modest compared to image-based ML problems; there is no evident need for `dimension reduction' such as principle component analysis.

\subsection{Random Forest Regressor}

The \textsc{Random Forest} algorithm is the optimal choice for this project due to its speed, accuracy, and flexibility, given the statistics of features on which it trains. It is based on a decision tree,  which can exactly predict the output values of the data. The drawback of any one decision tree is that it often tunes itself to the subtle details of the training set features. A decision tree can fail miserably when applied to similar but distinct test data; its focus on detail may miss the broad correlations that connect the features and the output. The \textsc{Random Forest} algorithm skirts this problem of `over-fitting' by creating multiple decision trees with arbitrary subsets of data. Combining these multiple decision trees allows broader correlations to emerge in the output.

Here we use the 16 features from EDR3 and DR2 as described in \S\ref{slr}, along with $\log(\chi^2)$ from the HGCA catalog as our output values. We divide the HGCA stars randomly in half to generate training and test sets. We then fit the training set with the {Python} {Sklearn} \textsc{Random Forest} algorithm. The resulting regressor (a method in the \texttt{sklearn} toolkit) predicts output from the test set. Ideally, the regressor would predict exactly the $\log(\chi^2)$ value for all stars in the training and test sets. In practice, there may not be enough information in the features we chose to predict whether a star is accelerating. The hope is that there is enough information to make predictions that can guide us to promising candidates.

\begin{figure}
\centerline{\includegraphics[width=4.5in]{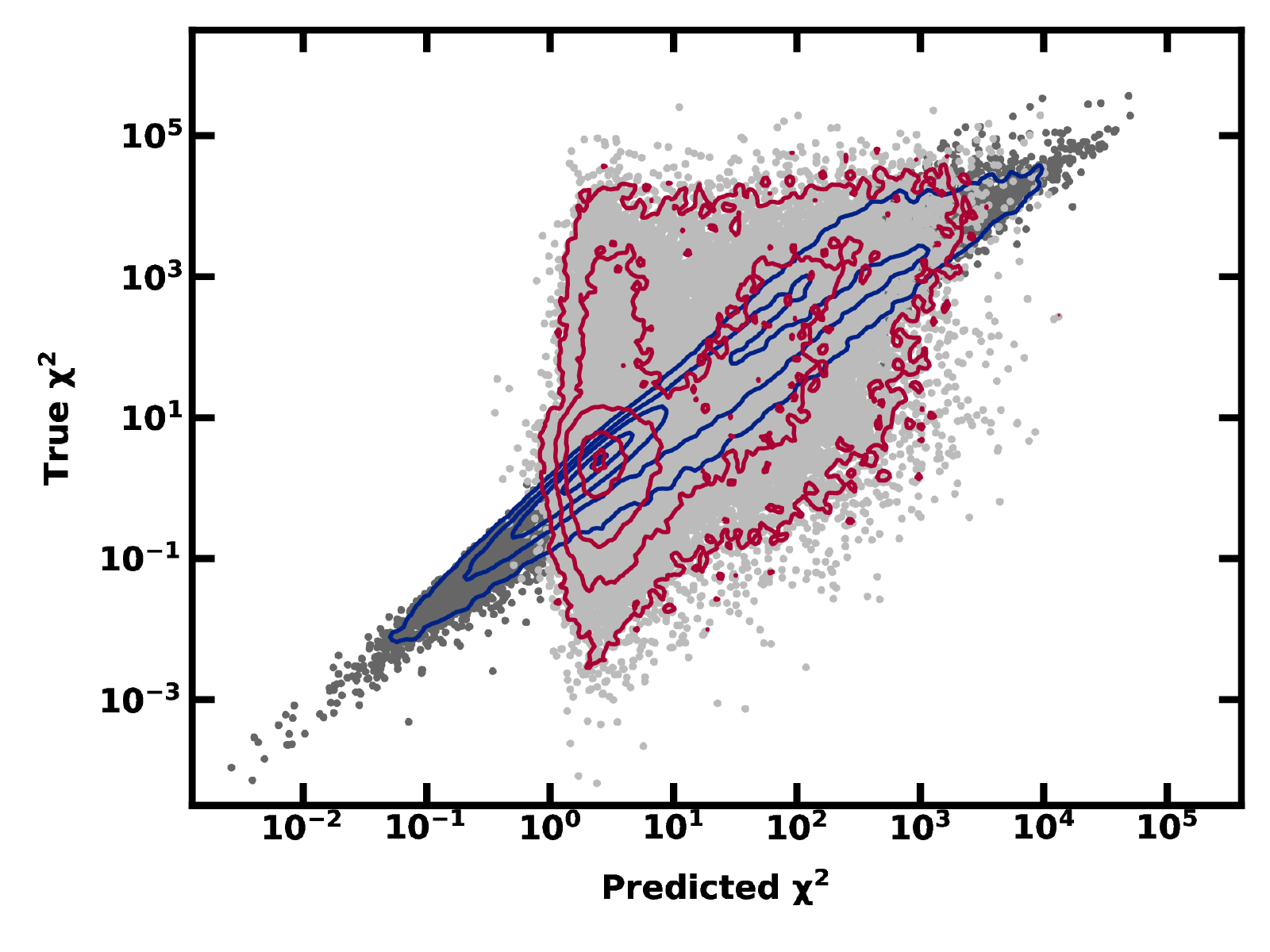}} % updated 023/01/21
    \caption{Predicted versus true $\chi^2$ for linear drift in the HGCA from \gaia\ DR2 and EDR3 astrometry. The true $\chi^2$ values are reported in the HGCA \citep{brandt2021}, while the predictions are generated with the \textsc{Random Forest} regressor. The darker points and blue contours (at containment thresholds of 2\%, 20\%, 50\%, 80\%, and 98\%) correspond to the training set, with half of the sources in the HGCA chosen at random. The remaining stars constitute the test set. The lighter-shaded points and red contours indicate the test set's predicted versus true $\chi^2$. The training set results are characteristic of the \textsc{Random Forest} regressor; it is derived from a decision tree algorithm, which gives perfect correlations between the true and predicted values. With \textsc{Random Forest}, this correlation is blurred by randomness in the decision tree branch choices, seen in the blue contours---they show a strong but imperfect correlation between true and predicted $\chi^2$. This approach preserves broader significant trends in the data, allowing them to be revealed in the test set. Without the randomness, the regressor would be overly tuned to the details of the training set, significantly reducing the effectiveness as a predictor of $\chi^2$ in other data. Here, the predicted $\chi^2$ shows a correlation between true values and predictions, which becomes stronger at larger $\chi^2$.
    \label{fig:contour}}
\end{figure}

Figure~\ref{fig:contour} illustrates the promise of the machine learning approach for predicting accelerating star candidates. While the regressor misses many stars that are accelerating with high significance --- some stars are predicted to have low values of $\chi^2$ even though their true $\chi^2$ values are quite high --- those stars with a high predicted $\chi^2$ tend to have high true values of $\chi^2$. Thus, while the predictor misses some fraction of suitable candidates, when it finds a candidate, that star is likely to be accelerating.

Table~\ref{tab:traintest} confirms this overall assessment. It illustrates that the machine learning approach yields candidates with an accuracy of over 80\%\ when we adopt stringent thresholds in predicted $\chi^2$ values. For example, in our training set, we can identify candidates with an 80\%\  likelihood of being a true `5-$\sigma$' accelerating star if we use a threshold of 500 for the predicted $\chi^2$ value. A negative consequence is that we identify only 7\%\ of the population of true accelerating stars. If we lower the threshold, our accuracy drops, but we identify a greater fraction of the true accelerating stars.

\begin{deluxetable}{c|ccc|ccc|ccc}
\tabletypesize{\footnotesize}
\tablecolumns{12}
\tablewidth{0pt}
\tablecaption{HGCA sources: Predictions of $\chi^2$ with and without DR2.
    \label{tab:traintest}}
\tablehead{
\colhead{\ } & \multicolumn{3}{c}{EDR3 $+$ DR2} & \multicolumn{3}{c}{EDR3 only}  & \multicolumn{3}{c}{low \ruwe\  test set}  
\vspace*{-5.5pt}\\
\colhead{$\chi^2$ threshold } & \colhead{$N_\text{pred}$} & \colhead{accuracy} & \colhead{efficiency} & \colhead{$N_\text{pred}$} & \colhead{accuracy} & \colhead{efficiency} & \colhead{$N_\text{pred}$} &  \colhead{accuracy} & \colhead{efficiency} 
}
\decimals
\startdata
% frame corrected bright only, updated 2023/01/22
 28.75 & 11268 & 67.9\% & 53.5\% &  11728 & 64.9\% & 53.3\% & 1482 & 60.4\% & 14.4\% \\
 50 & 8727 & 72.3\% & 44.2\% &  9234 & 67.8\% & 43.8\% & 967 & 67.3\% & 10.5\% \\
 100 & 5568 & 77.8\% & 30.3\% &  5271 & 70.9\% & 26.2\% & 575 & 75.0\% & 6.9\% \\
 250 & 2612 & 83.0\% & 15.2\% &  1608 & 73.8\% & 8.3\% & 274 & 81.0\% & 3.6\% \\
 500 & 1196 & 86.6\% & 7.3\% &  478 & 74.9\% & 2.5\% & 134 & 88.8\% & 1.9\% \\
 1000 & 491 & 87.2\% & 3.0\% &  108 & 71.3\% & 0.5\% & 39 & 97.4\% & 0.6\% \\
\enddata
\tablecomments{To generate the data contained here, HGCA sources listed in both DR2 and EDR3 were randomly divided in half to make the training and test sets. Training was performed both with and without DR2 data. The columns with the header $N_\text{pred}$ indicate the number of stars with a predicted $\chi^2$ greater than the threshold in the first column of each row. The accuracy indicates the likelihood that a star with a predicted $\chi^2$ above a threshold value (column 1) is ``unambiguously'' accelerating, with a true $\chi^2$ value above 28.75 (5 $\sigma$).  The efficiency refers to the fraction of all unambiguously accelerating stars ($\chi^2 \geq 28.75$) in the test set that are correctly identified using the predicted $\chi^2$ values above the threshold in column 1.} 
\end{deluxetable}

Table~\ref{tab:traintest} also illustrates the value of including the proper motion data from \gaia\ DR2 in the list of features we use to train the regressor. Without the DR2 data, we experience a decline in accuracy (the likelihood that a candidate is truly accelerating) and efficiency (a fraction of truly accelerating sources we identify as candidates).

Even with both EDR3 and DR2 parameters in the training data, the ML regressor is unable to achieve an accuracy above 90\%. In the best case with predicted $\chi^2$ values above 1000, over 10\%\ of the sources --- 63 out of 491 --- are false positives, objects that have a true $\chi^2$ below our 5-$\sigma$ threshold. To gain some insight as to why the regressor generated these false positives, we focus on the \ruwe\ parameter. Of the 63 false positives, 61 have \ruwe\ values above 1.3, suggesting binarity. Using the \simbad\ Astronomical Database, we confirm that 30 of these stars are associated with known binaries, including 20 objects identified as spectroscopic binaries, which are poorly fit with linear drift models \citep[e.g.,][]{stassun2021}. The ML regressor appears to be picking up these ``extra'' sources whose motion is not well matched to the acceleration sensitivity in the HGCA.

Finally, we consider how the regressor operates on a test set of sources with low \ruwe. Values of \ruwe\ below 1.3 indicate linear drift over the two-year duration of the \gaia\ DR3 observations, while high values of $\chi^2$ ($>28.75$) indicate acceleration over decades. The combination of low \ruwe\ and high $\chi^2$ among the accelerating candidates may signify stars with a lower mass companion or a more distant binary partner.  Table~\ref{tab:traintest} summarizes the regressor output on low-\ruwe\ stars. The efficiency in recovering true, unambiguously accelerating stars is lower for these candidates than for HGCA in general.  The accuracy, however, is comparable when the threshold for predicted $\chi^2$ is high ($\gtrsim$100). While low-\ruwe\ accelerating stars may be missed more than their high-\ruwe\ counterparts, the ML approach can still identify promising candidates on the basis of \gaia\ data alone.

\section{\catlongname}

We now apply our machine-learning algorithm to build a catalog of accelerating star candidates from \gaia~EDR3 data. The starting point is to select stars in EDR3 that have a parallax greater than 10~mas and are brighter than 17.5~mag in the Gaia B band. We keep stars in this group that are positively cross-matched with \gaia\ DR2 sources; we exclude stars also in the HGCA. These selection criteria yield a set of 226,943 stars. The parallax and brightness cuts limit the number of sources to nearby bright stars, increasing the likelihood that our catalog contains only stars with high-quality astrometric measurements. 

Next, we train the \textsc{Random Forest} regressor on the entire 115,160 sources in HGCA and make predictions of $\chi^2$ for each of the almost 227k stars from EDR3. By adopting a minimum predicted $\chi^2$ threshold of 28.75, corresponding to a 5-$\sigma$ outlier for the null hypothesis that a source has no acceleration, we define a new \catlongname\ (\catshortname), with \ncatstars\ stars. 
% An accompanying machine-readable table lists these stars, 
A machine-readable table accompanying the published article list these stars,
including their source identifiers, spectral information, and the list of features used in the machine-learning regression. Table~\ref{tab:cat} lists the top 25 sources in the \catshortname\ by $\chi^2$ value,  highlighting some of the most promising candidates in the \gaia\ database. The table also contains 25 stars in the catalog selected at random to provide a glimpse of more typical sources.

\begin{figure}
\centerline{\includegraphics[width=6.5in]{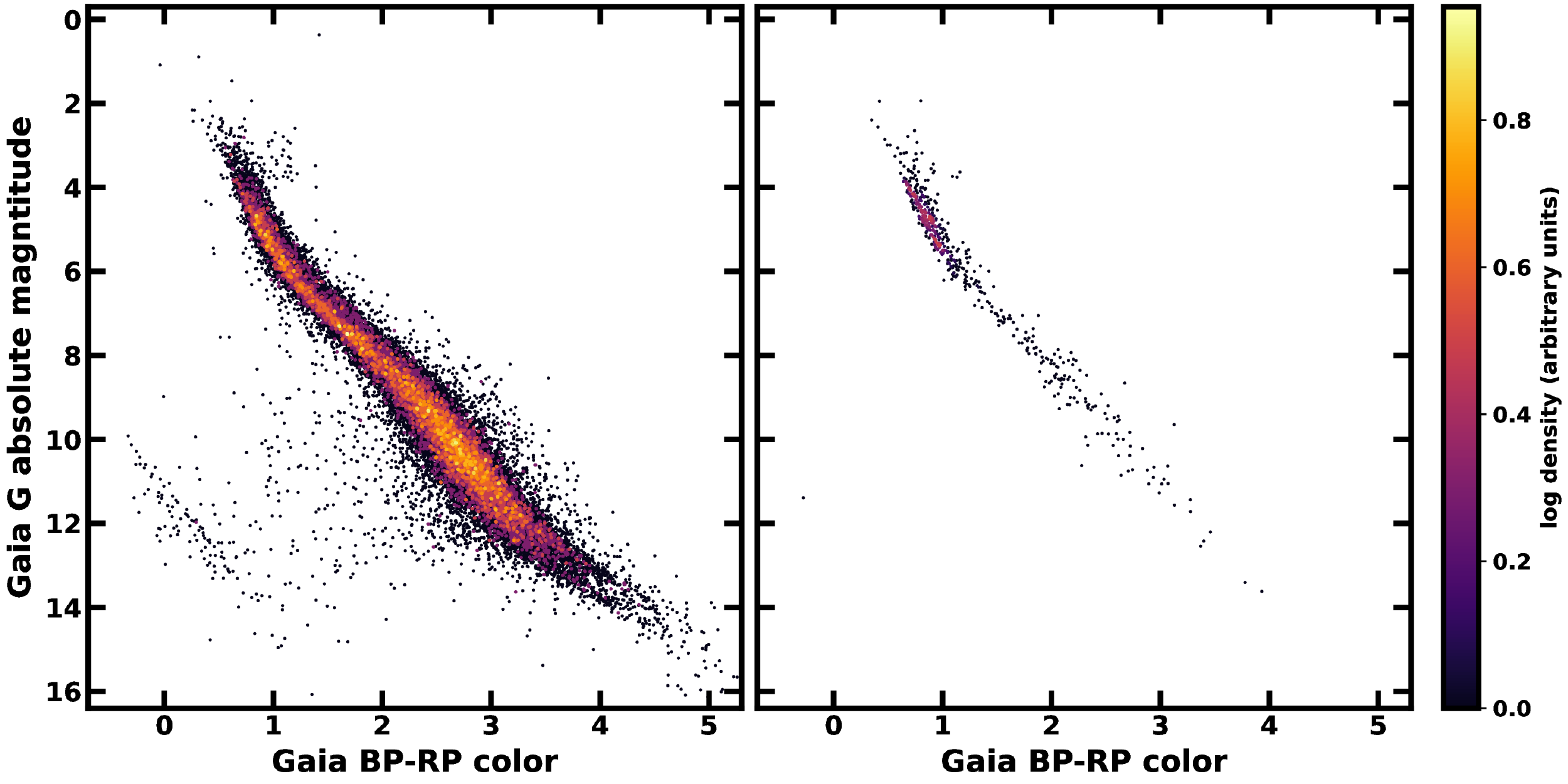} % updated with low density pdf 2023/01/22
}
%\vspace{-1.5in}
   \caption{The color-magnitude diagram of the \catlongname, derived from a machine-learning algorithm as described in the text. The left panel is the complete catalog containing \ncatstars\ stars from \gaia\ EDR3. Parallax and brightness cuts ensure that the stars included here are those with high-quality astrometric measurements. The panel on the right shows the CMD for our catalog's top 500 sources, which contain the subset of the most promising candidates sorted by $\chi^2$ values. The panel on the right is more heavily weighted toward the upper main sequence than the bulk of the sample (left), which has a more significant fraction of lower main sequence stars. The unique catalog of stars offers promising accelerating candidates intended to lead to the discovery of new stellar and substellar companions. 
    \label{fig:cat}}
\end{figure}

Figure~\ref{fig:cat} shows the full CMD of our results, illustrating that our sources are mainly late-type main-sequence stars. A companion plot shows that sources with the highest predicted $\chi^2$ are among the brighter stars, and many will turn out to be members of stellar binaries. A strong majority of stars in the \catshortname, about 95\%, have a \gaia\ EDR3 \ruwe\ parameter above the nominal binarity threshold of 1.3. The remaining 5\%, 1515 stars in our catalog, have \ruwe\ below this threshold. These objects may be excellent candidates as hosts of substellar companions, similar to Gl 229, HR 7672, Gl 758, HD 19467 and others high $\chi^2$ and low \ruwe \citep{brandt2019, brandtgm2021}.

\begin{deluxetable}{lrrrrrrrrrrrr}
\tabletypesize{\scriptsize}
\tablecolumns{12}
\tablewidth{0pt}
\tablecaption{\\\catlongname
    \label{tab:cat}}
\tablehead{
\colhead{\ }
&            \colhead{GAIA DR3 source\_id} &   \colhead{$\ell$} &           \colhead{b} &   \colhead{parallax} &          \colhead{pm} &  \colhead{G mag} &     \colhead{bp\_rp} &  
\colhead{ruwe} & \colhead{fmp} & \colhead{nss} & \colhead{nei} &  \colhead{chi2acc}
}
\decimals
\startdata
  1 & 1389522117550542464 &  64.114 &  53.781 &  44.333 & 444.648 & 7.203 & 1.050 & 2.271 &  44 & 0 & 1 & 5993.30 \\
  2 &  507836421987745536 & 131.565 &  -2.057 &  20.868 & 188.624 & 8.868 & 0.948 & 1.689 &   0 & 0 & 0 & 5026.56 \\
  3 & 3210018566592226816 & 207.138 & -22.935 &  19.073 & 160.670 & 8.393 & 0.896 & 3.014 &   7 & 0 & 0 & 3968.93 \\
  4 & 4445793563749783424 &  27.575 &  30.968 &  14.982 & 122.237 & 6.907 & 0.682 & 1.455 &  50 & 0 & 1 & 3850.37 \\
  5 & 5210494447648534400 & 289.325 & -22.055 &  21.921 & 176.882 & 9.041 & 1.128 & 2.485 &  73 & 0 & 1 & 3556.94 \\
  6 & 3168845665766691328 & 203.450 &  16.725 &  12.760 &  94.185 & 8.364 & 0.669 & 2.474 &   0 & 1 & 0 & 3019.87 \\
  7 & 5521657651369417216 & 264.627 &  -3.266 &  16.023 &  54.289 & 8.365 & 0.727 & 3.660 &   0 & 0 & 0 & 2972.77 \\
  8 & 2083800694741387904 &  84.898 &   5.742 &  15.968 &  39.610 & 9.022 & 0.857 & 2.719 &   0 & 1 & 0 & 2927.99 \\
  9 & 4498899219464393728 &  42.336 &  16.720 &  17.871 & 104.277 & 8.350 & 0.871 & 5.267 &   0 & 0 & 0 & 2849.18 \\
 10 & 1237090738916392832 &  23.087 &  61.357 & 148.179 & 225.687 & 6.430 & 1.518 & 1.510 &   0 & 0 & 1 & 2803.79 \\
 11 &  464640698944363264 & 137.768 &   1.364 &  13.285 &  36.481 & 8.431 & 0.686 & 1.769 &   0 & 3 & 0 & 2623.44 \\
 12 & 4894801780022148608 & 222.426 & -40.316 &  18.348 &  74.925 & 8.937 & 0.936 & 4.131 &   0 & 0 & 1 & 2545.69 \\
 13 &  543889717493219072 & 133.694 &  11.699 &  19.685 &  29.310 & 8.280 & 0.907 & 1.834 &  43 & 0 & 1 & 2523.82 \\
 14 & 2776209724185243392 & 119.057 & -49.908 &  21.007 &  93.617 & 9.224 & 1.124 & 1.660 &   0 & 0 & 0 & 2371.40 \\
 15 & 6336296021212109824 & 355.414 &  44.488 &  17.104 & 173.703 & 8.205 & 0.942 & 4.501 &   0 & 0 & 0 & 2361.74 \\
 16 &  767158369693232256 & 172.226 &  68.971 &  15.070 &  62.641 & 8.718 & 0.907 & 1.702 &   0 & 0 & 0 & 2314.92 \\
 17 & 3836011042119352576 & 237.554 &  42.620 &  10.110 &  51.667 & 8.828 & 0.696 & 2.968 &   0 & 1 & 0 & 2274.09 \\
 18 & 2974726131970829952 & 221.724 & -33.024 & 119.574 & 316.714 & 7.792 & 1.906 & 1.818 &  47 & 0 & 0 & 2261.79 \\
 19 & 3444211725808078464 & 179.296 &  -0.510 &  13.354 &  20.909 & 8.366 & 0.698 & 1.970 &   0 & 1 & 0 & 2257.14 \\
 20 & 5410307772352425600 & 271.773 &   1.824 &  10.319 &  39.308 & 7.787 & 0.469 & 1.645 &   0 & 0 & 0 & 2241.50 \\
 21 & 1990027642977047680 & 104.719 &  -5.709 &  10.633 &  76.348 & 8.477 & 0.922 & 2.129 &   1 & 1 & 0 & 2235.49 \\
 22 &  250294033837671296 & 150.156 &  -3.320 &  14.723 & 108.829 & 9.306 & 0.889 & 2.277 &   0 & 3 & 0 & 2234.09 \\
 23 & 2175444507873336448 &  94.221 &   2.325 &  12.142 &  60.938 & 7.506 & 0.766 & 2.295 &   0 & 0 & 0 & 2231.28 \\
 24 & 6792436799477051904 &  11.138 & -36.344 & 101.972 & 423.100 & 9.605 & 3.126 & 2.264 &   0 & 0 & 1 & 2208.51 \\
 25 & 4078101065647163904 &  10.502 &  -8.927 &  22.795 &  34.265 & 7.172 & 0.707 & 2.192 &   0 & 2 & 0 & 2191.14 \\
 2260 & 2160237059465812736 &  93.198 &  25.753 &  10.419 &  33.567 & 14.247 & 2.461 & 6.114 &   0 & 1 & 0 &  303.88 \\
 4231 & 3255963240506193024 & 191.902 & -32.729 &  11.705 &  49.832 & 13.096 & 2.247 & 6.201 &   0 & 0 & 0 &  218.89 \\
 4895 & 6095186723401124608 & 314.232 &  14.402 &  10.765 & 112.518 & 12.384 & 1.761 & 9.424 &   0 & 2 & 0 &  203.93 \\
 5147 & 1971859278487532416 &  88.205 &  -2.345 &  12.030 &  20.897 & 11.564 & 1.485 & 4.192 &   3 & 0 & 0 &  198.43 \\
 5566 & 6751641245892692992 &  10.488 & -25.186 &  19.483 & 147.308 & 14.511 & 2.802 & 13.535 &  50 & 0 & 0 &  190.69 \\
 9649 & 6013511293139455360 & 334.828 &  16.501 &  10.448 &  23.271 & 14.716 & 2.671 & 7.982 &   0 & 0 & 0 &  137.11 \\
 12118 & 2349851722325492864 &  88.587 & -84.503 &  10.511 &  21.385 & 14.217 & 2.469 & 2.409 &  14 & 0 & 1 &  115.36 \\
 12271 & 6409644648059644544 & 332.100 & -46.629 &  12.420 & 138.498 & 15.856 & 2.591 & 4.840 &  51 & 0 & 0 &  114.11 \\
 12983 & 6560454693719133952 & 348.767 & -52.234 &  10.695 &  68.793 & 10.718 & 1.049 & 2.900 &   0 & 2 & 0 &  108.71 \\
 13681 & 5615274503760936704 & 238.983 &  -2.507 &  16.657 &  62.507 & 9.692 & 1.060 & 1.173 &  18 & 0 & 0 &  103.40 \\
 15852 & 3761210823700893440 & 260.317 &  42.425 &  13.367 & 114.473 & 14.150 & 2.256 & 2.584 &   0 & 0 & 0 &   88.76 \\
 16581 & 2718461174471202560 &  82.679 & -41.811 &  16.947 & 175.124 & 15.321 & 2.996 & 1.367 &  70 & 0 & 1 &   84.16 \\
 16849 & 6410224399925039872 & 333.807 & -44.564 &  19.677 &  39.287 & 14.504 & 2.775 & 3.600 &   0 & 0 & 0 &   82.65 \\
 17893 & 2277592402263269120 & 111.314 &  17.894 &  11.358 & 173.663 & 15.132 & 2.472 & 1.369 &  35 & 0 & 1 &   76.53 \\
 18178 & 6289950850187390848 & 326.798 &  39.330 &  11.598 & 128.002 & 9.416 & 0.873 & 1.590 &   0 & 0 & 0 &   75.11 \\
 18562 & 1372466046503024000 &  59.067 &  48.803 &  17.026 & 163.693 & 16.069 & 3.242 & 2.318 &   0 & 1 & 0 &   73.23 \\
 20056 & 4043958072852457984 & 358.763 &  -3.626 &  10.150 & 135.790 & 15.475 & 2.876 & 2.082 &   0 & 0 & 0 &   65.93 \\
 20684 &  810519363081689216 & 169.347 &  53.743 &  12.869 & 138.491 & 15.843 & 3.272 & 1.775 &   0 & 0 & 0 &   63.06 \\
 22443 & 6491462331939518080 & 325.436 & -52.476 &  13.053 &  66.407 & 16.578 & 3.692 & 1.442 &   0 & 0 & 0 &   55.45 \\
 22504 & 5248874554583059584 & 282.275 &  -9.722 &  15.886 & 304.587 & 13.993 & 2.701 & 7.904 &   0 & 0 & 0 &   55.22 \\
 23925 & 1824638794717734656 &  56.330 &  -3.113 &  11.417 & 116.092 & 16.893 & 2.448 & 1.171 &   4 & 0 & 0 &   49.44 \\
 24661 & 4790404967733742720 & 249.571 & -40.817 &  18.516 & 169.585 & 17.469 & 4.169 & 1.363 &   0 & 0 & 0 &   46.53 \\
 25062 & 4499162243258532224 &  37.705 &  19.330 &  13.596 &  36.872 & 13.859 & 2.461 & 15.790 &  24 & 0 & 1 &   44.99 \\
 26156 & 3644327824224945408 & 334.737 &  53.261 &  13.233 & 152.183 & 16.029 & 2.993 & 1.301 &   0 & 0 & 0 &   40.96 \\
 28951 & 1661565719739103488 & 109.500 &  53.878 &  11.029 &  36.710 & 15.893 & 2.607 & 1.545 &  25 & 0 & 0 &   31.04 \\
 \enddata
\tablecomments{This list is the top 25 stars in the \catshortname, 
sorted by decreasing $\chi^2$ (`chi2acc'), along with 25 randomly selected sources from the catalog. 
The units of sky position ('$\ell$' and 'b', Galactic coordinates) are degrees, 
parallax is in mas, while the `pm' field is proper motion in mas/year, 
the 'ruwe' label refers to the DR3 renormalized unit-weighted error, and 'fmp' is DR3's \texttt{ipd\_frac\_multi\_peak} parameter, giving the percent of image fields associated with each source in which multiple peaks are measured (an indicator of binarity). 
The column `nss' indicates whether a source is in the Gaia DR3 
\texttt{nss\_accelaration\_astro} table (1), \texttt{nss\_twobody\_orbit} table (2), both (3) or neither (0). 
The column labelled 'nei' is a Boolean flag indicating whether a source has a neighbor within a 
projected distance of 200~au. Some neighbors have measured parallax consistent with a source distance at the 5-$\sigma$ level, 
while the rest have no measured parallax. 
}
\end{deluxetable}

Our method and the \catlongname\ stars demonstrate a machine learning approach's performance in identifying accelerating star candidates with course-grained astrometric parameters.  With the release of DR3 and the non-single star (NSS) tables \texttt{nss\_acceleration\_astro} and \texttt{nss\_twobody\_orbit} \citep{halbwachs2022}, we have an opportunity to compare the \catshortname~candidates with accelerating sources that have been confirmed through detailed analysis of low-level \gaia\ data. Both Table~\ref{tab:cat} and the full machine-readable table have a column that indicates if a source is in one, both, or none of the \gaia~NSS tables. Of the top 25 stars, five are in \texttt{nss\_acceleration\_astro}, one is in \texttt{nss\_twobody\_orbit}, and an additional two sources are in both tables. The remaining 16 sources are not in either NSS table.

Table~\ref{tab:compnss} gives a broader overview of the overlap between our candidate accelerating stars and sources identified as non-single stars in DR3. The \catshortname\ contains over 97\%\ of the \texttt{nss\_acceleration\_astro} members and about {40}\%\ of the \texttt{nss\_twobody\_orbit} table when both \gaia\ tables are filtered with the same parallax and brightness cuts as in our catalog. The fraction of NSS stars recovered decreases as we increase the predicted $\chi^2$ threshold within the \catshortname. While the fraction of stars in our catalog that are also in NSS tables increases with increasing $\chi^2$ (over 34\%\ of the sources with $\chi^2 > 1000$ are in the NSS tables, compared to 24\%\ for the whole catalog) there is not a tight correspondence between our high $\chi^2$ values and \gaia\ NSS membership. This situation is expected; the training and test of the ML algorithm show that the predicted and true $\chi^2$ are not highly correlated. Furthermore, the \texttt{nss\_acceleration\_astro} table contains only stars where acceleration is identified at very high significance (the norm of the acceleration or its derivative, if measured, divided by the uncertainty is greater than 25). This cut selects only a subset of the potential candidates.

\begin{deluxetable}{r|rc|rc|rc}
\tabletypesize{\footnotesize}
\tablecolumns{7} % ????
\tablewidth{0pt}
\tablecaption{Comparison with DR3 non-single star sources.
    \label{tab:compnss}}
\tablehead{
\colhead{\ } & \multicolumn{2}{c}{this catalog$^\text{*}$} & \multicolumn{2}{c}{{nss\_acceleration\_astro}}
& \multicolumn{2}{c}{{nss\_two\_body\_orbit}} 
\vspace*{-5.5pt}\\
\colhead{$\chi^2$ threshold} & \colhead{\#} & \colhead{\% nss} & 
\colhead{\# in table} & \colhead{\% of table} & 
\colhead{\# in table} & \colhead{\% of table} 
}
\decimals
\startdata
% thresh, Ncat, Ncat/Ntot, Ncatacc, Ncatacc/Nacc, Ncat2bdy Ncat2bdy/N2bdy
% updated 2023/01/22 with bright-only frame corrections...
11.8 & 43524 & 16.8\% & 3866 & 99.3\% & 3528 & 43.6\% \\  
28.75 & 29684 & 23.4\% & 3802 & 97.6\% & 3209 & 39.7\% \\  
 100 & 14163 & 31.2\% & 2598 & 66.7\% & 1870 & 23.1\% \\  
 250 & 3276 & 34.0\% & 685 & 17.6\% & 454 &  5.6\% \\  
 500 & 845 & 34.9\% & 205 &  5.3\% & 102 &  1.3\% \\  
1000 & 215 & 34.4\% & 51 &  1.3\% & 27 &  0.3\% 
\enddata
\tablecomments{The data in this catalog are listed in DR2 and EDR3 with parallax greater than 10~mas and apparent magnitude less than 17.5 in the \gaia\ G band. HGCA sources are excluded. The threshold indicates the minimum value of the ML-predicted $\chi^2$ for the null hypothesis of linear drift.
The second column is the number of sources in the \catshortname\ above the threshold, while the third column is the fraction of these sources that are in either of the \texttt{nss\_acceleration\_astro} or \texttt{nss\_two\_body\_orbit} tables. The columns under the nss table headings are the number of objects above the given threshold, and the percentage of the total number of nss table sources that fit our input catalog's selection criteria. The fraction of objects in the two\_body orbit tables picked up in our analysis is lower presumably because the table includes close-in binaries that would not be identified by a HGCA-trained regressor.}
\end{deluxetable}

\mbox{}\vspace{22pt}\subsection{Sources with neighbors in \gaia}

In addition to cross-matches with the NSS tables, the \catshortname\ includes flags indicating whether a star has a neighbor in \gaia\ DR3 within a projected distance of 200~\au\ or less.  We identify neighbors with known parallaxes with \gaia\ archive queries for objects that have parallaxes consistent to within 5-$\sigma$; this generous threshold accounts for the possibility that \gaia\ DR3 parallax errors are underestimated, especially in binary systems \citep[e.g.,][]{chulkoc2022}. We also identify neighbors in \gaia\ DR3 within a projected distance of 200~\au\ that have unknown parallaxes. This subset of catalog members with identifiable companions allows us to explore the interplay between $\chi^2$ and companion characteristics, and the connection with \ruwe, which is a known indicator of binarity.

\begin{figure}
\centerline{
\includegraphics[width=6.5in]{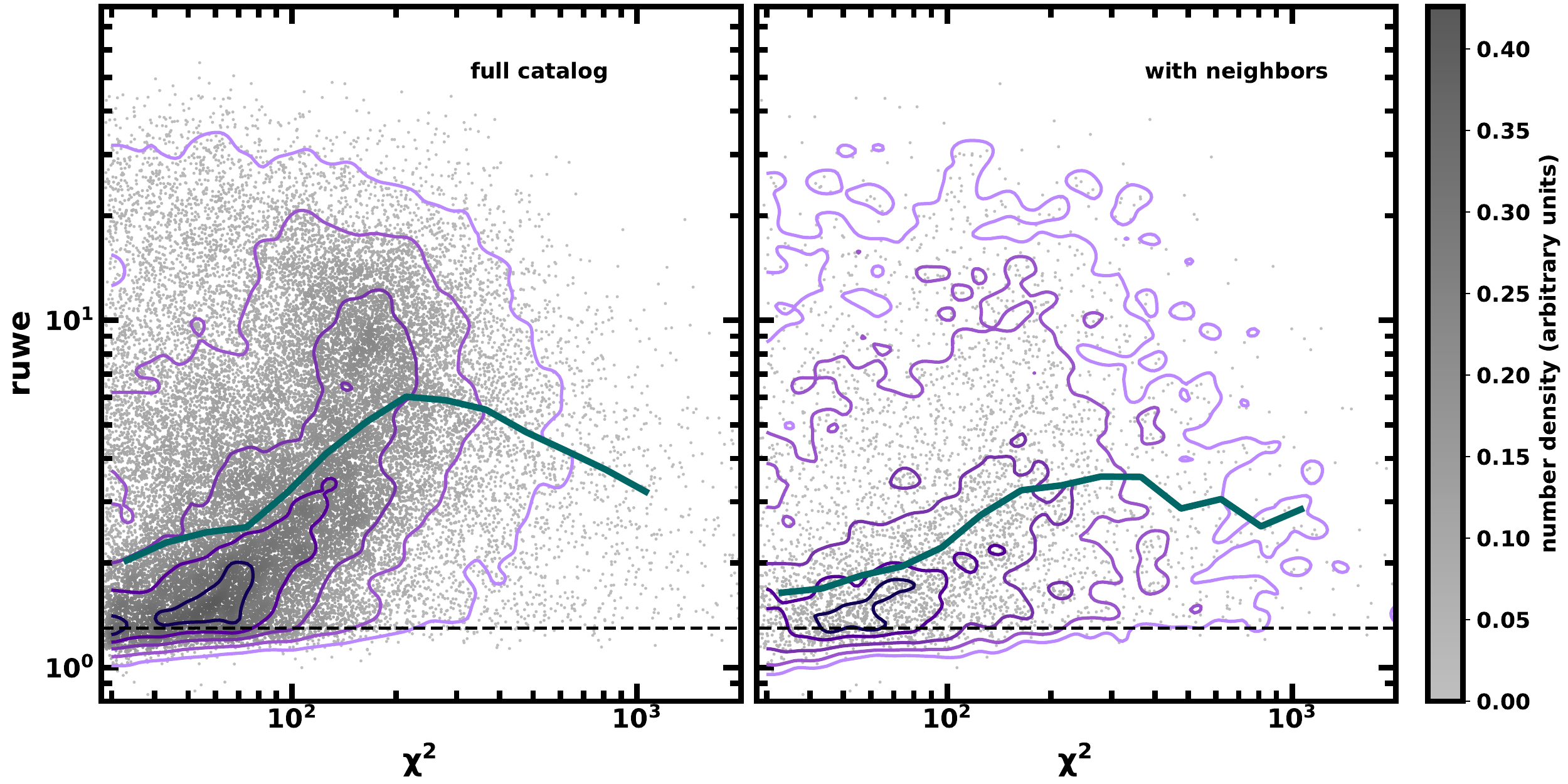} % updated 2023/01/22
}
   \caption{The renormalized unit-weighted error (\texttt{ruwe}) of sources in the \catshortname\ versus the predicted $\chi^2$ testing the null hypothesis for stellar acceleration. The horizontal dashed line is at a value of $\text{\texttt{ruwe}}= 1.3$, which nominally delimits binary source (above the line) and single sources (below) in the \gaia\ data.  The left and right panels are for the full catalog and only those stars with a neighbor inside of 200~\au\ identified in the \gaia\ archive. Contours are at containment thresholds of 5\%, 20\%, 50\%, 80\%, and 95\%.
    \label{fig:ruwecat}}
\end{figure}

Figure~\ref{fig:ruwecat} shows the location of sources with neighbors in the plane of \ruwe\ versus the predicted $\chi^2$, along with a similar plot with all stars in the catalog. As compared to the HGCA data in Figure~\ref{fig:ruwehgca}, the distribution of stars with neighbors has a higher median \ruwe. Similarly, the median \ruwe\ trends higher for sources with binary companions than for all \catshortname\ stars. The presence of detectable neighbors within the projected separation of 200~\au\ indicates that these stars tend to have higher \ruwe\ values, to stellar binaries \citep[e.g.,][]{brandt2018, belokurov2020, kervella2022, pearce2022}.

A separate \gaia\ DR3 parameter, \texttt{ipd\_frac\_multi\_peak}, provides an independent indication of whether a star has a neighbor. Its value is the fraction (as a percent) of images of a given source that have other sources (``multiple peaks'') in the same field of view. About 35\%\ of stars in the \catshortname\ have multi-peak detections, compared with just under 20\%\ of all stars in DR3 that meet our parallax and brightness cuts . While the overlap between NSS table membership and multi-peak detections among \catshortname\ stars is not strong (about 10\%\ of the NSS table members show non-zero \texttt{ipd\_frac\_multi\_peak}), 80\%\ of sources with potential neighbors within 200~\au\ have a non-zero \texttt{ipd\_frac\_multi\_peak} parameter. Furthermore, over half of the sources with multi-peak detections, 6380 stars, are not in either the NSS tables or our list of stars with potential neighbors inside of 200~\au. Thus, multi-peak detections may be a distinct flag for the presence of companions of \catshortname\ stars.  While we do not exploit this potential here, Table~\ref{tab:cat} and the 
% accompanying machine-readable table 
machine-readable table accompanying the published article
include \texttt{ipd\_frac\_multi\_peak} for reference.

\begin{figure}
\centerline{
\includegraphics[width=4.5in]{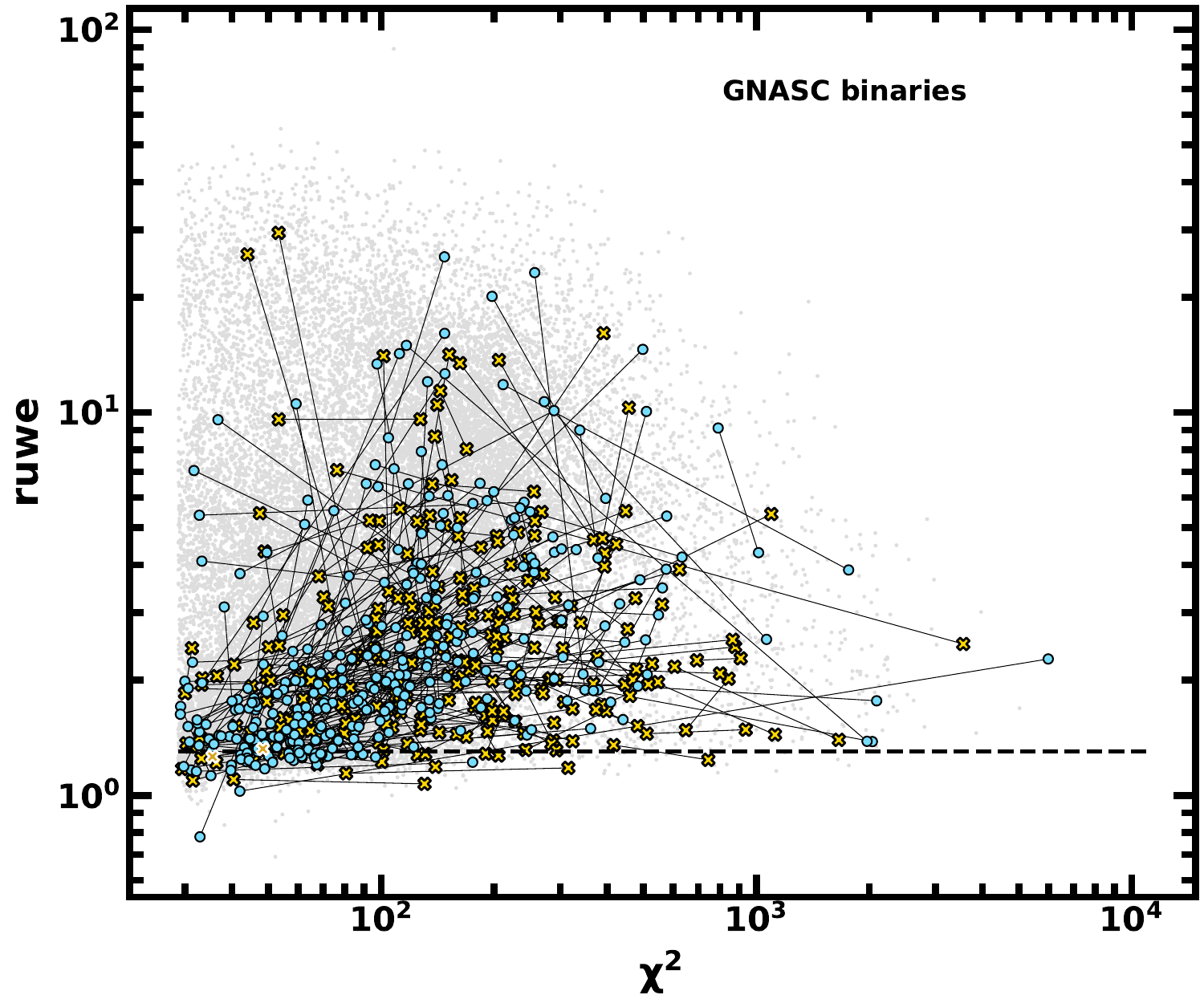} % updated 2023/01/22
}
   \caption{The renormalized unit-weighted error versus predicted $\chi^2$ for binaries in the \catshortname, similar to Fig.~\ref{fig:ruwecat}. The gold 'x' symbols mark the 139 binaries that appear in the \simbad\  Astronomical Database, while the light cyan circles are the 172 new binaries in our catalog. Each pair is connected with a thin line. The overall distributions of both known and new binaries are similar to each other and to the underlying distribution of all sources in the catalog (light gray points in the background). Notably, 7 pairs lie below the 1.3 \ruwe\ threshold for binarity.
    \label{fig:bin}}
\end{figure}

%% bcb binaries --> resolved binaries
A subset of stars in the \catshortname\ have neighbors that are also catalog members. These pairs constitute known binary systems as well as new candidates with projected separations within 200~\au. A total of 139 pairs are listed in the \simbad\ Astronomical Database as resolved binaries. Another 172 pairs of stars in the \catshortname\ are in the catalog but not flagged as binary partners in \simbad. These known and new stars appear in Figure~\ref{fig:bin}, showing their location in a plot of \ruwe\ versus $\chi^2$. While most of these sources have $\text{ruwe}\geq 1.3$, 57 out of 622 individual stars lie below that threshold. Seven binaries have both member stars with \ruwe below the threshold; two of these systems are listed in \simbad.

\subsection{\catshortname~Candidates of Interest}

Targeted searches on stars showing statistically significant gravitational pull have proven helpful in detecting unresolved astrometric binaries and alerting the presence of substellar companions. The machine learning algorithm's ability to seek out unique stars within the \gaia~data led us to many known double or multiple star systems, offering proof-of-concept for our method. One source, AT~Mic~B ({\small{\gaia\ EDR3 6792436799477051904}}, also GJ~799~B), is in the binary system and part of a hierarchical triple with AU~Mic \citep{2014AJ....147...85R}. AT~Mic~B is only slightly smaller than its close companion, AT~Mic~A; both are pre-main sequence M dwarfs with masses of approximately a quarter of a solar mass \citep{mccarthy2012}. With a separation of 2.6~arcsec at a heliocentric distance of about 10~pc, the orbital separation of the binary is about 25~au. From equation~(\ref{eq:dmu}), we estimate a change in the proper motion of the order of 5~mas/yr over a three-year time frame, roughly the window of observation spanned by DR3. This value is well within the limit of detectability.  

While AT~Mic~B has the 24th highest $\chi^2$ value in our catalog (Table~\ref{tab:cat}), it is not in either of the \gaia\ NSS tables.  Other high-ranking stars are similar; the star with the highest predicted $\chi^2$, for example, is HD 139341B (\small{\gaia\ DR3~1389522117550542464}), a K dwarf about 22 pc from us. Yet it has a  binary companion (\small{\gaia\ DR3~1389522117550541056}, also in our catalog) at a projected distance of $\sim$27~au and a possible additional companion at about 2700~au in projection (\small{\gaia\ DR3~1389522392427772288}, not in the \catshortname) with similar parallax and proper motion.

We noticed a known eclipsing binary while adjusting and modifying parameters ({\small{\gaia\ DR3 4439783323958670080}}). This source, BD+07~3142, was part of an analysis of multicolored light and radial velocities, where the authors \citep{2011A&A...525A..66D} determined that BD+07 3142 is a close over-contact interacting binary system, W subtype of W UMa. These low-mass binaries evolve into contact through orbital shrinking \citet{stepien2011} and knowing the initial masses is a direct link to understating their formation and evolution. Light-curve synthesis makes analyses of these systems using observations from smaller ground-based instruments possible due to their short orbital periods \citep{Olivera2021}. Since the individual stars are not resolvable by Gaia, BD+07~3142 is not expected to show astrometric evidence of acceleration despite its rapid binary motion. \citet{lohr2015} provided evidence for a third stellar partner based on period changes; our analysis, with the identification of BD+07~3142 as an accelerating star candidate, is also consistent with the presence of a more distant companion.

One candidate, ({\small{\gaia\ DR3 4692276170589724288}}), that stood out early did not make the final adjustment. Due to our strict selection criteria, we offer this source as a cautionary example of an excluded instance. This source is the primary in the HD 7693 binary, consisting of two K-type stars that are part of a stable 2 + 2 hierarchical quadruple system \citep{Fuhrmann2017}. The study of such hierarchical stellar systems provides information and insight into the role of dynamical and dissipative processes of star formation \citep{ Tokovinin2017}. Other sources within the catalog may prove challenging to determine their usefulness when they are situated in relatively active and busy regions of space. Parameter adjustments were made to reduce the number of crowded field sources. \cite{refId0} discusses limitations in detecting sources in overcrowded areas, resulting in partially overlapping windows. Such issues are said to have since been addressed with the newest release of DR3; however, a more profound analysis of crowded fields may not be possible until future data releases.

Finally, we note that all three members of a triple-star system, PM~J20324$+$2608, are in the \catshortname\ (\texttt{source\_id}s 1856618021650243328, 1856618021664457216, and 1856618021664457344); none of these sources appear in the \gaia\ non-single star tables. Another source group that potentially constitutes a multiple star system includes three members of the \catshortname, with \texttt{source\_id}s 4480062901652623232, 4480062867305057280, and 4480062828640260736. The first two stars are red dwarfs about an arcsecond apart; the third, also a late-type star, is about 30~arcsec away. It has a potential binary partner at about 5~arcsec separation, \texttt{source\_id}~4480062832945811584, that is not in the catalog. All four stars are, within errors, at the same heliocentric distance, approximately 48~pc. At that distance, the separation between these two binary pairs is roughly 1500~au. None of these sources are in the non-single star tables from DR3 or were linked to known sources in \simbad. If confirmed with more detailed observations, this 2$+$2 quadruple star might join other recent discoveries of multiple systems of red dwarfs \citep{vrijmoet2020, vrijmoet2022}

\section{Conclusion}

We have demonstrated the effectiveness of a machine-learning approach to identify nearby accelerating stars. The method is based on the idea that stellar acceleration yields proper motion anomalies evident in poor astrometric fits with data models that assume constant drift across the sky \citep{brandt2018, belokurov2020, kervella2022}. We take advantage of the Hipparcos-Gaia Catalog of Accelerations \citep{brandt2021} to train a regression algorithm to identify stars in \gaia~EDR3 that experience changes in measurable proper motion over decades. We focus on a $\chi^2$ metric that indicates how well an individual star's proper motion adheres to a linear-drift model \citep{brandt2021}; our machine-learning approach is derived from regression with this quantity, not on a binary label (acceleration vs. non-acceleration) to add flexibility to the interpretation of the predictions. With a \textsc{Random Forest} regressor, our algorithm predictions are for 226,943 stars in \gaia~EDR3, all within 100 pc of the Sun and brighter than 17.5 mag in Gaia~G~band. The result is the \catlongname\ (\catshortname), which contains \ncatstars\ stars with $\chi^2 > 28.75$, formally corresponding to a 5-$\sigma$ or better detection of acceleration.

The test-train phase of development suggests that the algorithm presented here identifies candidates with a likelihood of true acceleration ranging from 67\% at our threshold $\chi^2$ of 28.75 to over 85\%\ accuracy at high values of $\chi^2 \gtrsim 250$ (Table~\ref{tab:traintest}. The top stars in the \catshortname, ranked by decreasing values of $\chi^2$, are dominated by nearby members of stellar binaries.

This work began before the release of \gaia~DR3. Our catalog was generated from DR2 and EDR3 data with no explicit information about stellar acceleration. With its wealth of information, especially about non-single stars \citep{halbwachs2022}, the new DR3 catalog offers a perfect opportunity to assess our algorithm. The \catshortname\ contains over 97\%\ of the DR3 \texttt{nss\_acceleration\_astro} table and 40\%\ of the \texttt{nss\_twobody\_orbit} table. About a third of our top 25 candidates (Table~\ref{tab:cat}) are in these two tables, with most belonging to \texttt{nss\_acceleration\_astro}. This distribution makes sense; our algorithm trained on the HGCA is designed to look for proper motion anomalies over a long baseline, making it more sensitive to the slower astrometric accelerations modeled in the \texttt{nss\_acceleration\_astro} table.

Combined with the candidates for accelerating stars within \gaia-DR3, our catalog provides a comprehensive set of potential stellar and substellar companions within 100~pc of the Sun. Recent deep imaging observations have shown considerable promise in identifying companions that would be missed in traditional radial velocity and imaging surveys \citep[e.g.,][]{currie2020, sutlieff2021, derosa2023}. Future studies based on these catalogs will enable a better understanding of the architectures of stellar binaries and planetary systems.

\section*{Acknowledgement}
We thank Prof.~Gail Zasowski for advice and interpretion of the CMD and stars in our catalog, and Prof.~Dan Wik for his input and discussions on stellar astrophysics. We are grateful to the referee, Dr.~Timothy Brandt, for provided comments and guidance that significantly improved this work. MW and JH acknowledge support from the University of Utah through the Undergraduate Research Opportunity Program. This work has made use of data from the European Space Agency (ESA) mission
{\it Gaia} (\url{https://www.cosmos.esa.int/gaia}), processed by the {\it Gaia}
Data Processing and Analysis Consortium (DPAC,
\url{https://www.cosmos.esa.int/web/gaia/dpac/consortium}). Funding for the DPAC has been provided by national institutions, in particular the institutions
participating in the {\it Gaia} Multilateral Agreement. This research has also made use of the Aladin sky atlas, developed at CDS, Strasbourg Observatory, France, and the SIMBAD Astronomical Database, operated at CDS, Strasbourg, France.

%\bibliography{stars}{}
\bibliography{main.bbl}{}

\end{document}